\documentstyle[12pt]{article}\textwidth 16.5cm
\begin{document}
%
%%%%%%%%%%%%%%%%%%%%%%%%%%%%%%%%%%%%%%%%%%%%%%%%%%%%%%%%%%%%%%
%
\begin{titlepage}
\begin{raggedleft}
THES-TP 96/05\\
March 1996\\
\end{raggedleft}
\vspace{2em}
\begin{center}
{\Large\bf{ BFV analysis of the $U_{EM}(1)$ gauged SU(3) WZW \\
\vspace{0.25em}
 model and the Faddeev-Jackiw approach}}
\footnote {Work supported by the European Community Human
Mobility program "Eurodafne",Contract CHRX-CT92-00026.}\\
\vspace{2em}
{\large J.E.Paschalis and P.I.Porfyriadis}
\footnote{On leave from the Tbilisi State University, Tbilisi,
Georgia.}
\vspace{1em}\\
Department of Theoretical Physics, University of
Thessaloniki,\\GR 54006 \hspace{0.25em},\hspace{0.25em} Thessaloniki
\hspace{0.25em},\hspace{0.25em} Greece\\
\vspace{1em}
{\scriptsize PASCHALIS@OLYMP.CCF.AUTH.GR\\
PORFYRIADIS@OLYMP.CCF.AUTH.GR}
\end{center}
\vspace{2em}
\begin{abstract}
The four dimensional SU(3) WZW model coupled to electromagnetism is 
treated as a constrained system in the context of Batalin-Fradkin-Vilkovisky 
formalism. It is shown that this treatment is equivalent to the 
Faddeev-Jackiw (FJ) approach. It is also shown that the field 
redefinitions that transform the fields of the model into BRST and 
$\sigma$ closed are actually the Darboux's transformations used in 
the FJ formalism.
\end{abstract}
\end{titlepage}
%
%%%%%%%%%%%%%%%%%%%%%%%%%%%%%%%%%%%%%%%%%%%%%%%%%%%%%%%%%%%%%%%%%
%
\section{Introduction}
\label{intr}
In \cite{PP3} the SU(2) WZW model coupled to electromagnetism was
treated as a constrained system in the context of the
Batalin-Fradkin-Vilkovisky (BFV) formalism \cite{BFV}. 
Comparison was made
between this method and the Faddeev-Jackiw (FJ) approach to
constrained systems. Common features were emphasized. In this
letter we extend this analysis in the SU(3) case.

In the FJ \cite{F-J} approach we start with a Lagrangian
density first order in time derivatives. Then by using Darboux's
theorem and Euler-Lagrange equations we transform it into an
expression whose canonical one-form is diagonal and where the
constraints occur in specific terms linear in coordinate
variables. Next we solve the constraint equations and after 
substituting the solutions into the expression for the Lagrangian 
density we repeat
the whole process again until we end up with an unconstrained
Lagrangian density, with diagonal canonical one form and whose phase
space is reduced. On the contrary in the BFV formalism the phase
space of the theory is extended. This is done in two steps.
First a canonical momentum conjugate to every Lagrange
multiplier is introduced (which has to vanish) so increasing the
number of constraints. Second a ghost field is introduced for
every constraint so extending the phase space of the theory. The
gauge fixing is done by properly choosing the gauge fermion.
%%%%%%%%%%%%%%%%%%%%%%%%%%%%%%%%%%%%%%%%%%%%%%%%%%%%%%%%%%%%%%%%%%
\section{The SU(3) WZW model coupled to electromagnetism}
\label{model}
In \cite{PP2} the SU(3) WZW model coupled to electromagnetism 
\cite{Witten,Balachandran} was treated as a constrained system 
in the context of the FJ formalism. This model 
describes the low energy interactions of the eight Goldstone 
bosons and photons including those related to the axial anomaly.  
The effective action of the model up to second order in the pion
fields $\theta_{a}\;,\; a=1,...,8 \;$ is given by
\begin{eqnarray}
  {\cal L}_{eff}&\!\!\!\!=&\!\!\!\!{\cal L}_{EM}+
                 {\cal L}_{\sigma}^{(2)}+{\cal L}_{WZW}^{(2)}
                 +O(\theta^3)\; \; ,\\
  {\cal L}_{EM}&\!\!\!\!=&\!\!\!\!-\frac{1}{4}
                F_{\mu \nu}F^{\mu \nu}\; \; ,\nonumber \\
  {\cal L}_{\sigma}^{(2)}&\!\!\!\!=&\!\!\!\! \frac{1}{2}
               (\partial_\mu \theta_{a})(\partial^\mu\theta_{a})+
                eA^\mu(\theta_{2}\partial_\mu\theta_{1}-
                \theta_{1}\partial_\mu\theta_{2}+
                \theta_{5}\partial_\mu\theta_{4}-
                \theta_{4}\partial_\mu\theta_{5}) \nonumber \\
      &\!\!\!\!+&\!\!\!\! \frac{e^2}{2}A_\mu A^\mu
          (\theta_{1}^{2}+\theta_{2}^{2}+\theta_{4}^{2}+\theta_{5}^{2}) 
                        \; \; ,\nonumber \\
  {\cal L}_{WZW}^{(2)}&\!\!\!\!=&\!\!\!\!-\frac{N_{c}e^2}{12\pi^{2}f_\pi}
               {\epsilon^{\mu \nu \alpha \beta}}A_\mu(\partial_\nu
               A_\alpha)(\partial_\beta \theta_{3}+\frac{1}{\sqrt{3}} 
                         \partial_\beta\theta_{8}) \; \; .\nonumber 
\end{eqnarray}
and as an expression first order in time derivatives of the fields is  
written as follows
\vspace{1em}
\begin{eqnarray}
{\cal L}_{eff}&\!\!\!\!=&\!\!\!\!-\mbox{\boldmath $\pi$}\cdot\dot{\bf A}\rm+
                 p_{\it a}\dot{\theta_{\it a}}-
                 H^{(2)}_0
                -A_{0}(\rho^{(2)}-\bf\nabla\cdot\mbox{\boldmath $\pi$}\rm)
                +O(\theta^3) \;\; , \\
\nonumber \\
 H^{(2)}_0 &\!\!\!\!=&\!\!\!\!
\frac{1}{2}[\mbox{\boldmath $\pi$}^{\rm 2}+\bf B^{\rm 2}
              +(\nabla \rm\theta_{\it a})^{2}+p_{\it a}^{2}]
              +\frac{e^2}{2}\bf A^{\rm 2}\rm
          (\theta_{1}^{2}+\theta_{2}^{2}+\theta_{4}^{2}+\theta_{5}^{2})
                 \nonumber \\
         &\!\!\!\!+&\!\!\!\!
              e\bf A\cdot\rm(\theta_{1}\bf \nabla \rm \theta_{2}-
                              \theta_2 \bf \nabla \rm \theta_1+
                              \theta_{4}\bf \nabla \rm \theta_{5}- 
              \theta_5 \bf \nabla \rm \theta_4) -
              \frac{N_{c}e^2}{6\pi^2 f_\pi} 
   (\mbox{\boldmath $\pi$}\cdot \bf B\rm)
(\theta_{3}+\frac{1}{\sqrt{3}}\theta_{8})
                     \; ,\nonumber \\
\nonumber \\
\rho^{(2)}&\!\!\!\!=&\!\!\!\! e(p_{2}\theta_1-p_{1}\theta_{2}+ 
               p_{5}\theta_4-p_{4}\theta_{5}) \; \; .\nonumber 
\end{eqnarray}
\vspace{0.5em}
See Appendix for notation.

The corresponding BFV action is given by
\begin{equation}
S_{BFV}=\int d^{4} x (-\mbox{\boldmath $\pi$} \cdot \dot{\bf A} \rm +
\pi_{0} \dot{A}_{0} 
+ p_{\it a}\dot{\theta}_{\it a} + \dot{C}{\cal P} + 
\dot{\bar{C}}\bar{{\cal P}}
- H_{0}^{(2)}) + \int d t [\Psi,Q] )
\end{equation}
The scalar potential has become a full dynamical variable and
its conjugate momentum $\pi_0$ has to vanish.
We have also introduced the canonical pair 
$(C,{\cal P})$ of a ghost field and its conjugate mometum 
corresponding to the constraint
$ G_{1}=\rho^{(2)}-\nabla\cdot\mbox{\boldmath $\pi$}$,
and the canonical pair $ (\bar{C},\bar{{\cal P}}) $ of an
antighost field and its canonical momentum, corresponding to
the constraint 
$ G_{2}=\pi_{0} $. $\Psi$ is the gauge fermion and Q is the BRST
charge. The two constraints are first class.
 The BRST charge is given by 
\begin{equation}
Q=\int d^{3}x [C(\rho^{(2)} - \nabla \cdot \mbox{\boldmath $\pi$}
\rm) +  i \bar{{\cal P}} \pi_{0}]
\end{equation}
The canonical Hamiltonian $\int d^3 x H^{(2)}_0$ and $ S_{BFV} $
are invariant under the BRST transformations
\begin{displaymath}
  s\bf A\rm=-\nabla C\;\; , \;\; \hspace{1em}
      sC=0\;\; , \;\;
\end{displaymath}
\begin{displaymath}
s{\cal P}=\nabla\cdot\mbox{\boldmath $\pi$} -\rho^{(2)}\;\; , \;\;
\hspace{1em} 
      s\mbox{\boldmath $\pi$}=0\;\; , \;\;
\end{displaymath}
\begin{displaymath}
  sA_{0}=i\bar{{\cal P}}\;\; , \;\; \hspace{1em}
      s\bar{{\cal P}}=0\;\; , \;\;
\end{displaymath}
\begin{equation}
  s\bar{C}=-i\pi_{0}\;\; , \;\; \hspace{1em}
      s\pi_{0}=0\;\; , \;\;
\end{equation}
\begin{displaymath}
  s\theta_{1}=-e\theta_{2}C\;\; , \;\; \hspace{1em}
      s\theta_{2}=e\theta_{1}C\;\; , \;\;
\end{displaymath}
\begin{displaymath}
  sp_{1}=-ep_{2}C\;\; , \;\; \hspace{1em}
      sp_{2}=ep_{1}C\;\; , \;\;
\end{displaymath}
\begin{displaymath}
  s\theta_{4}=-e\theta_{5}C\;\; , \;\; \hspace{1em}
      s\theta_{5}=e\theta_{4}C\;\; , \;\;
\end{displaymath}
\begin{displaymath}
  sp_{4}=-ep_{5}C\;\; , \;\; \hspace{1em}
      sp_{5}=ep_{4}C\;\; , \;\;
\end{displaymath}
\begin{displaymath}
  s\theta_{3}=s\theta_{6}=s\theta_{7}=s\theta_{8}=0\;\; , \;\;
\end{displaymath}
\begin{displaymath} 
      sp_{3}=sp_{6}=sp_{7}=sp_{8}=0\;\; . \;\;
\end{displaymath}

We decompose $\bf A\rm$ and $\mbox{\boldmath $\pi$}$ into 
transverse $(\bf A^{\rm T}\; , \;\mbox{\boldmath $\pi$}^{\rm T})$ 
and longitudinal $(\bf A^{\rm L}\; ,\; \mbox{\boldmath $\pi$}^{\rm L})$
componenents
\begin{displaymath}
 \bf A^{\rm T}=\bf A\rm -\nabla A^{L'}
             \;\; , \;\; \hspace{1em}
     \bf A^{\rm L}\rm=\nabla A^{L'}
            \;\; , \;\; \hspace{1em}
      A^{\rm L'}\rm=\frac{1}{\nabla^2}
(\nabla \cdot \bf A)\rm\;\; , \;\;
\end{displaymath}
\begin{displaymath}
\mbox{\boldmath $\pi$}^{T}=\mbox{\boldmath $\pi$}
-\frac{\nabla}{\nabla^2} \pi^{L'} \;\; , \;\; \hspace{1em}
\mbox{\boldmath $\pi$}^{L}=\frac{\nabla}{\nabla^2} \pi^{L'}\;\;
, \;\; 
\hspace{1em}
      \pi^{L'}=
\nabla \cdot \mbox{\boldmath $\pi$}\rm\;\; . \;\;
\end{displaymath}

The following field redefinitions
\begin{displaymath}
   p_1\rightarrow p_1 \cos{\alpha}+p_2\sin{\alpha} 
           \;\; , \;\; \hspace{1em}
   \theta_1 \rightarrow \theta_1 \cos{\alpha}+\theta_2 \sin{\alpha}
            \;\; , \;\;
\end{displaymath}
\begin{displaymath}
   p_2 \rightarrow p_2 \cos{\alpha}-p_1\sin{\alpha}
           \;\; , \;\; \hspace{1em}
   \theta_2 \rightarrow \theta_2 \cos{\alpha}-\theta_1 \sin{\alpha}
           \;\; , \;\;
\end{displaymath}
\begin{equation}
   p_4\rightarrow p_4 \cos{\alpha}+p_5\sin{\alpha} 
           \;\; , \;\; \hspace{1em}
   \theta_4 \rightarrow \theta_4 \cos{\alpha}+\theta_5 \sin{\alpha}
           \;\; , \;\;
\end{equation}
\begin{displaymath}
   p_5 \rightarrow p_5 \cos{\alpha}-p_4\sin{\alpha}
           \;\; , \;\; \hspace{1em}
   \theta_5 \rightarrow \theta_5 \cos{\alpha}-\theta_4 \sin{\alpha}
           \;\; , \;\;
\end{displaymath}
where $\alpha = eA^{L'}$,
transform $\theta_i$ and $p_i$ fields $(i=
1,2,4,5)$ into BRST and $\sigma$ closed (physical). So we
have
\begin{displaymath}
  sA^{L'}\rm=-C\;\; , \;\; \hspace{1em}
      sC=0\;\; , \;\;
\end{displaymath}
\begin{displaymath}
s{\cal P}=\pi^{L'}-\rho^{(2)}\;\; , \;\;
\hspace{1em} 
      s\pi^{L'}=0\;\; , \;\;
\end{displaymath}
\begin{displaymath}
  sA_{0}=i\bar{{\cal P}}\;\; , \;\; \hspace{1em}
      s\bar{{\cal P}}=0\;\; , \;\;
\end{displaymath}
\begin{equation}
  s\bar{C}=-i\pi_{0}\;\; , \;\; \hspace{1em}
      s\pi_{0}=0\;\; , \;\;
\end{equation}
\begin{displaymath}
  s\theta_{\it a}=0\;\; , \;\; \hspace{1em}
      sp_{\it a}=0\;\; , \;\; \hspace{1em}
\it a=\rm 1,2,...,8
\end{displaymath}
\begin{displaymath}
  s\bf A^{\rm T}\rm=0\;\; , \;\; \hspace{1em}
      s\mbox{\boldmath $\pi$}^T=0\;\; , \;\;
\end{displaymath}
and
\begin{displaymath}
  \sigma(-C)=A^{L'}\;\; , \;\; \hspace{1em}
      \sigma A^{L'}=0\;\; , \;\;
\end{displaymath}
\begin{displaymath}
\sigma \pi^{L'}={\cal P}\;\; ,
\;\; \hspace{1em} 
      \sigma {\cal P}=0\;\; , \;\;
\end{displaymath}
\begin{displaymath}
  \sigma(i\bar{{\cal P}})=A_{0}\;\; , \;\; \hspace{1em}
      \sigma A_{0}=0\;\; , \;\;
\end{displaymath}
\begin{equation}
  \sigma(-i\pi_{0})=\bar{C}\;\; , \;\; \hspace{1em}
      \sigma(\bar{C})=0\;\; , \;\;
\end{equation}
\begin{displaymath}
  \sigma\theta_{\it a}=0\;\; , \;\; \hspace{1em}
      \sigma p_{\it a}=0\;\; , \;\; \hspace{1em}
\it a =\rm 1,2,...,8
\end{displaymath}
\begin{displaymath}
  \sigma \bf A^{\rm T}\rm=0\;\; , \;\; \hspace{1em}
      \sigma \mbox{\boldmath $\pi$}^{T}=0\;\; , \;\;
\end{displaymath}
where $\sigma$ is the contracting homotopy operator \cite{Henneaux}. 
Note that there is no way that we can transform the fields 
$\bf A^{\rm L}\rm, A_0, C, \bar{C}, {\cal P}\rm, \bar{{\cal P}}$ into
BRST and $\sigma$ closed.

Now we solve for C, $\bar{{\cal P}}$ ,
$ \mbox{\boldmath $\pi$}^L $ and $\pi_0$ from (7) 
\begin{equation}
C=-sA^{L'}\;\; , \;\;
\hspace{1em} 
\bar{{\cal P}}=-isA_{0} \;\; , \;\; \hspace{1em}
\mbox{\boldmath $\pi$}^L
=\frac{\nabla}{\nabla^2}(s{\cal P}+\rho^{(2)})\;\; , 
\;\; \hspace{1em}
\pi_0=is\bar{C}\;\; ,
\end{equation}
and after performing the transformations (6) in (3) we substitute
$ C , \bar{\cal P}, \mbox{\boldmath $\pi$}^L, \pi_0$ from (9). We
end up with the following expression for $S_{BFV}$
\begin{eqnarray}
S_{BFV}&\!\!\!\!\rightarrow&\!\!\!\!\int d^4 x[-\mbox{\boldmath $\pi$}^T
\cdot\dot{\bf A}^{\rm T}+ 
             \rm p_{\it a}\dot{\theta_{\it a}} - H_{C}^{(2)}+sF^{(2)}]
             +\int dt [\Psi,Q]   
     \; \; ,\\
   H_{C}^{(2)}
          &\!\!\!\!=&\!\!\!\!\frac{1}{2}[(\mbox{\boldmath $\pi$}^T)^2
+\bf B^{\rm 2}\rm-
                  \rho^{(2)}\frac{1}{\bf \nabla \rm^2}
                  \rho^{(2)}+(\bf \nabla \rm \rm\theta_{\it a})^{2}
                  +p_{\it a}^{2}]\nonumber \\
          &\!\!\!\!+&\!\!\!\! e\bf A^{\rm T}\cdot\rm
                 (\theta_{1} \bf \nabla \rm \theta_{2}-
                  \theta_2 \bf \nabla \rm \theta_1+
                  \theta_{4} \bf \nabla \rm \theta_{5}-
                  \theta_5 \bf \nabla \rm \theta_4)
                 + \frac{e^2}{2}(\bf A^{\rm T}\rm)^2
                    (\theta_{1}^{2}+\theta_{2}^{2}+
                    \theta_{4}^{2}+\theta_{5}^{2}) \nonumber \\
          &\!\!\!\!-&\!\!\!\!\frac{N_{c}e^2}{6\pi^2 f_\pi}
                [\mbox{\boldmath $\pi$}^T +
                \frac{\bf \nabla \rm}{\bf \nabla \rm^2}\rm \rho^{(2)}] 
                \cdot\bf B\rm(\theta_3+
                \frac{1}{\sqrt{3}}\theta_8) \;\; , \nonumber \\
  F^{(2)} &\!\!\!\!=&\!\!\!\!
  i\bar{C}\dot{A}_{0}+
{\cal P}\dot{A}^{L'}+\frac{1}{2}(s{\cal P})\frac{1}{\nabla^2}{\cal P}+
{\cal P}\frac{1}{\nabla^2}\rho^{(2)}+\frac{N_{c}e^2}{6\pi^2 f_\pi}
(\frac{\nabla}{\nabla^2}{\cal P})\cdot \bf B\rm(\theta_3+
\frac{1}{\sqrt{3}}\theta_8) \nonumber \;\;,
\end{eqnarray}
where $H_{C}^{(2)}$ is the Coulomb gauge Hamiltonian.

Finally we fix the gauge fermion as follows
\begin{displaymath}
  \Psi=-\int d^3 x F^{(2)}
\end{displaymath}
and we get a Coulomb gauge expression for the effective
action with the unphysical $\bf A^{\rm L}$ cancelled out.
In \cite{PP2} in the context of the FJ formalism we solved
the constraint $ \rho^{(2)}-\nabla\cdot\mbox{\boldmath $\pi$}=0$
 for $\mbox{\boldmath $\pi$}^L$ and we
substituted back into the expression (2) for the effective
Lagrangian density. We came up with an uncanonical
expression which was diagonalized by performing the 
transformations (6). So it turns out that (6) are actually the Darboux's
transformations of the FJ formalism needed for this case.
%
%%%%%%%%%%%%%%%%%%%%%%%%%%%%%%%%%%%%%%%%%%%%%%%%%%%%%%%%%%%%%%%%%%
%
\section{Keeping next order terms}
\label{exp-third}

The effective Lagrangian density up to third order in $\theta_{\it a}$ 
is given by
\begin{equation}
    {\cal L}_{eff}={\cal L}_{EM}+{\cal L}_{\sigma}^{(2)}+
                   {\cal L}_{WZW}^{(2)}+{\cal L}_{WZW}^{(3)}
                   +O(\theta^4)\; \; ,
\end{equation}
where  ${\cal L}_{\sigma}^{(2)}$ and
${\cal L}_{WZW}^{(2)}$ are given in (1) and
\begin{eqnarray}
   {\cal L}_{WZW}^{(3)}&\!\!\!\!=&\!\!\!\!
               -\frac{N_{c}e}{3\pi^{2}f_{\pi}^3}
               {\epsilon^{\mu \nu \alpha \beta}}(\partial_\mu A_\nu)
              (\theta_1\partial_\alpha \theta_2 -
                \theta_2 \partial_\alpha \theta_1+
                \theta_4\partial_\alpha \theta_5 -
                \theta_5 \partial_\alpha \theta_4)
               (\partial_\beta \theta_3+\frac{1}{\sqrt{3}}
                 \partial_\beta \theta_8) \nonumber \\
   &\!\!\!\!-&\!\!\!\! \frac{N_{c}e}{\sqrt{3}\pi^{2}f_{\pi}^3}
               {\epsilon^{\mu \nu \alpha \beta}}(\partial_\mu A_\nu)
               (\theta_7\partial_\alpha \theta_6 -
                \theta_6 \partial_\alpha \theta_7)\partial_\beta
                \theta_8 \nonumber \\
        &\!\!\!\!+&\!\!\!\! \frac{N_{c}e^2}{18\pi^{2}f_{\pi}^3}
                {\epsilon^{\mu \nu \alpha \beta}}(\partial_\mu A_\nu)
                      (\partial_\alpha A_\beta)
               \{ [4(\theta_{1}^2+\theta_{2}^2)+
                5(\theta_{4}^2+\theta_{5}^2)] \theta_3 \nonumber \\ 
        &\!\!\!\!+&\!\!\!\!
             \sqrt{3}[2(\theta_{1}^2+\theta_{2}^2)+
                \theta_{4}^2+\theta_{5}^2]\theta_8 +
         2[(\theta_1 \theta_5 - \theta_2 \theta_4)\theta_7 + 
          (\theta_1 \theta_4 + \theta_2 \theta_5)\theta_6] \}
                    \nonumber \\
        &\!\!\!\!-&\!\!\!\! \frac{N_{c}e^2}{3\pi^{2}f_{\pi}^3}
                {\epsilon^{\mu \nu \alpha \beta}}
                A_\mu (\partial_\nu A_\alpha)
                (\theta_3 + \frac{1}{\sqrt{3}}\theta_{8})
                \partial_\beta (\theta_{1}^2+\theta_{2}^2 +
                \theta_{4}^2+\theta_{5}^2) \nonumber  \; \; .
\end{eqnarray}
In the non-covariant notation, as an expression first order in time 
derivatives of the fields, ${\cal L}_{eff}$ is given by 
\vspace{1em}
\begin{equation}
   {\cal L}_{eff}=-\mbox{\boldmath $\pi$}\cdot \dot{\bf A\rm}\rm +
                  p_{\it a}\dot{\theta_{\it a}}
            - H^{(2)}_0 -H^{(3)}_0 -A_{0}(\rho^{(2)}+\rho^{(3)}
            -\bf\nabla\cdot\mbox{\boldmath $\pi$}\rm)+O(\theta^4)\; \; ,
\vspace{1em}
\end{equation}
\vspace{1em}
where  $H^{(2)}_0$ and $\rho^{(2)}$ are given in (2) and
\begin{eqnarray}
\vspace{1em}
   H^{(3)}_0 &\!\!\!\!=&\!\!\!\!-\frac{N_{c}e}{3\pi^{2}f_{\pi}^3}
              [\mbox{\boldmath $\pi$}\times\bf \nabla \rm (\theta_3 + 
              \frac{1}{\sqrt{3}} \theta_8)-
              (p_3+\frac{1}{\sqrt{3}}p_8) \bf B\rm)]
              \cdot(\theta_1\bf \nabla \rm \theta_2-\theta_2\bf \nabla
              \rm\theta_1+\theta_4\bf \nabla \rm \theta_5-
              \theta_5\bf \nabla \rm\theta_4) \nonumber \\
          &\!\!\!\!-&\!\!\!\!\frac{N_{c}e}{\sqrt{3}\pi^{2}f_{\pi}^3}
              (\mbox{\boldmath $\pi$}
\times\bf \nabla \rm \theta_8-p_8 \bf B\rm)
              \cdot(\theta_7\bf \nabla \rm \theta_6-\theta_6\bf \nabla
              \rm\theta_7) \nonumber \\
      &\!\!\!\!+&\!\!\!\! \frac{N_{c}e^2}{9\pi^{2}f_{\pi}^3}
              (\mbox{\boldmath $\pi$}\cdot \bf B\rm)
              \{ [4(\theta_{1}^2+\theta_{2}^2)+
                5(\theta_{4}^2+\theta_{5}^2)] \theta_3 \nonumber \\ 
        &\!\!\!\!+&\!\!\!\!
             \sqrt{3}[2(\theta_{1}^2+\theta_{2}^2)+
                \theta_{4}^2+\theta_{5}^2]\theta_8 +
         2[(\theta_1 \theta_5 - \theta_2 \theta_4)\theta_7 + 
          (\theta_1 \theta_4 + \theta_2 \theta_5)\theta_6] \}
                    \nonumber \\
      &\!\!\!\!-&\!\!\!\! \frac{N_{c}e^2}{3\pi^{2}f_{\pi}^3}
              (\mbox{\boldmath $\pi$}\times \bf A\rm) \cdot [\bf \nabla
          \rm(\theta_{1}^2+\theta_{2}^2+\theta_{4}^2+\theta_{5}^2)] 
             (\theta_3+\frac{1}{\sqrt{3}}\theta_8)\nonumber \\
      &\!\!\!\!-&\!\!\!\! \frac{2N_{c}e^2}{3\pi^{2}f_{\pi}^3}
            (\bf A\cdot B\rm)(p_1\theta_1+p_2\theta_2+p_4\theta_4+ 
            p_5\theta_5)(\theta_3+\frac{1}{\sqrt{3}}\theta_8)\nonumber \\
       &\!\!\!\!-&\!\!\!\! \frac{N_{c}e}{3\pi^{2}f_{\pi}^3}
           [\bf B\cdot\bf \nabla \rm(\theta_3+\frac{1}{\sqrt{3}}\theta_8)]
           (p_2\theta_1-p_1\theta_2+p_5\theta_4-p_4\theta_5)\nonumber \\
       &\!\!\!\!-&\!\!\!\! \frac{N_{c}e}{\sqrt{3}\pi^{2}f_{\pi}^3}
           (\bf B\cdot\bf \nabla \rm \theta_8)
           (p_6\theta_7-p_7\theta_6) \; \; , \nonumber \\
\nonumber \\
\nonumber \\
  \rho^{(3)}&\!\!\!\!=&\!\!\!\!-\frac{N_{c}e^2}{3\pi^{2}f_{\pi}^3}
                      \bf \nabla \rm\cdot[\bf B\rm
              (\theta_{1}^2+\theta_{2}^2+\theta_{4}^2+\theta_{5}^2)
              (\theta_3+\frac{1}{\sqrt{3}}\theta_8)] \; \; .\nonumber 
\end{eqnarray}

The BFV effective action is given by 
\begin{equation}
S_{BFV}=\int d^{4} x (-\mbox{\boldmath $\pi$} \cdot \dot{\bf A} \rm +
\pi_{0} \dot{A}_{0} 
+ p_{\it a}\dot{\theta}_{\it a} + \dot{C}{\cal P} + 
\dot{\bar{C}}\bar{{\cal P}}
- H_{0}^{(2)}- H_{0}^{(3)}) + \int dt [\Psi,Q] 
\end{equation}
As in the previous case $\pi_0$ is the canonical momentum
 conjugate to $ A_0 $ and has to vanish and
we have also introduced the canonical pair 
$(C,{\cal P})$  corresponding to the constraint
$ G_{1}=\rho^{(2)}+\rho^{(3)}-\nabla\cdot\mbox{\boldmath $\pi$}$,
and the canonical pair $ (\bar{C},\bar{{\cal P}}) $
 corresponding to the constraint $ G_{2}=\pi_{0} $.
Also by $\Psi$ and Q we denote the gauge 
 fermion and the BRST charge respectively.
\begin{equation}
Q=\int d^{3}x [C(\rho^{(2)}+\rho^{(3)} - 
\nabla \cdot \mbox{\boldmath $\pi$} \rm) +  i \bar{{\cal P}} \pi_{0}]
\end{equation}

The canonical Hamiltonian $\int d^3 x (H^{(2)}_0+H^{(3)}_0)$ 
and $ S_{BFV} $ are invariant under the following BRST transformations
\begin{displaymath}
  s\bf A\rm=-\nabla C\;\; , \;\; \hspace{1em}
      sC=0\;\; , \;\;
\end{displaymath}
\begin{displaymath}
s{\cal P}=\nabla\cdot\mbox{\boldmath $\pi$}
-\rho^{(2)}-\rho^{(3)} \;\; , \;\;
\hspace{1em} 
s\mbox{\boldmath $\pi$}=\frac{N_{c}e^2}{3\pi^{2}f_{\pi}^3}
\nabla[(\theta_{1}^2+\theta_{2}^2+\theta_{4}^2+\theta_{5}^2)
(\theta_3+\frac{1}{\sqrt{3}}\theta_8)]\times\nabla C 
\;\; , \;\;
\end{displaymath}
\begin{displaymath}
  sA_{0}=i\bar{{\cal P}}\;\; , \;\; \hspace{1em}
      s\bar{{\cal P}}=0\;\; , \;\;
\end{displaymath}
\begin{equation}
  s\bar{C}=-i\pi_{0}\;\; , \;\; \hspace{1em}
      s\pi_{0}=0\;\; , \;\;
\end{equation}
\begin{displaymath}
  s\theta_{1}=-e\theta_{2}C\;\; , \;\; \hspace{1em}
 sp_{1}=-ep_{2}C-\frac{2N_{c}e^2}{3\pi^{2}f_{\pi}^3}
(\bf B\rm\cdot\nabla C) \theta_1 (\theta_3+\frac{1}{\sqrt{3}}\theta_8)
      \;\; , \;\;
\end{displaymath}
\begin{displaymath}
  s\theta_{2}=e\theta_{1}C
\;\; , \;\; \hspace{1em}
      sp_{2}=ep_{1}C-\frac{2N_{c}e^2}{3\pi^{2}f_{\pi}^3} 
(\bf B\rm\cdot\nabla C) \theta_2 (\theta_3+\frac{1}{\sqrt{3}}\theta_8)
\;\; , \;\;
\end{displaymath}
\begin{displaymath}
  s\theta_{3}=0\;\; , \;\; \hspace{1em}
      sp_{3}=-\frac{N_{c}e^2}{3\pi^{2}f_{\pi}^3}
(\bf B\rm\cdot\nabla C) (\theta_{1}^2+ \theta_{2}^2+
\theta_{4}^2+ \theta_{5}^2)
\;\; , \;\;
\end{displaymath}
\begin{displaymath}
  s\theta_{4}=-e\theta_{5}C\;\; , \;\; \hspace{1em}
 sp_{4}=-ep_{5}C-\frac{2N_{c}e^2}{3\pi^{2}f_{\pi}^3}
(\bf B\rm\cdot\nabla C) \theta_4 (\theta_3+\frac{1}{\sqrt{3}}\theta_8)
      \;\; , \;\;
\end{displaymath}
\begin{displaymath}
  s\theta_{5}=e\theta_{4}C
\;\; , \;\; \hspace{1em}
      sp_{5}=ep_{4}C-\frac{2N_{c}e^2}{3\pi^{2}f_{\pi}^3} 
(\bf B\rm\cdot\nabla C) \theta_5 (\theta_3+\frac{1}{\sqrt{3}}\theta_8)
\;\; , \;\;
\end{displaymath}
\begin{displaymath}
  s\theta_{8}=0\;\; , \;\; \hspace{1em}
      sp_{8}=-\frac{N_{c}e^2}{3\sqrt{3}\pi^{2}f_{\pi}^3}
(\bf B\rm\cdot\nabla C) (\theta_{1}^2+ \theta_{2}^2+
\theta_{4}^2+ \theta_{5}^2)
\;\; , \;\;
\end{displaymath}
\begin{displaymath}
s\theta_{6}=s\theta_{7}=sp_{6}=sp_{7}=0 \;\; . \;\;
\end{displaymath}
The following field redefinitions
\begin{equation}
 \begin{array}{l}
\vspace{0.5em}
\theta_1 \rightarrow \theta_1 \cos{\alpha}+\theta_2 \sin{\alpha}
\;\; , \;\; \\
\vspace{0.5em}
\theta_2 \rightarrow \theta_2 \cos{\alpha}-\theta_1 \sin{\alpha}
\;\; , \;\; \\
\vspace{0.5em}
\theta_4 \rightarrow \theta_4 \cos{\alpha}+\theta_5 \sin{\alpha}
\;\; , \;\; \\
\vspace{0.5em}
\theta_5 \rightarrow \theta_5 \cos{\alpha}-\theta_4 \sin{\alpha}
\;\; , \;\; \\
\vspace{0.5em}
\mbox{\boldmath $\pi$}^T\rightarrow \mbox{\boldmath $\pi$}^T-
\frac{N_{c}e^2}{3\pi^{2}f_{\pi}^3} 
\bf \nabla \rm[(\theta_{1}^2+\theta_{2}^2+
\theta_{4}^2+\theta_{5}^2)(\theta_3+\frac{1}{\sqrt{3}}\theta_8)]\bf 
                        \times A^{\rm L}\;\;,\\
\vspace{0.5em}
p_1\rightarrow p_1 \cos{\alpha}+p_2\sin{\alpha}
+\frac{2N_{c}e^2}{3\pi^{2}f_{\pi}^3} 
               (\bf B\cdot A^{\rm L}\rm)
(\theta_1 \cos{\alpha}+\theta_2 \sin{\alpha})
(\theta_3+\frac{1}{\sqrt{3}}\theta_8)\;\;,\\ 
\vspace{0.5em}
p_2 \rightarrow p_2 \cos{\alpha}-p_1\sin{\alpha}
+\frac{2N_{c}e^2}{3\pi^{2}f_{\pi}^3} 
               (\bf B\cdot A^{\rm L}\rm) 
(\theta_2 \cos{\alpha}-\theta_1 \sin{\alpha})
(\theta_3+\frac{1}{\sqrt{3}}\theta_8)\;\;,\\ 
\vspace{0.5em}
  p_3 \rightarrow p_3 +\frac{N_{c}e^2}{3\pi^{2}f_{\pi}^3}
               (\bf B\cdot A^{\rm L}\rm)(\theta_{1}^2+
                 \theta_{2}^2+\theta_{4}^2+
                 \theta_{5}^2)\;\; , \\
\vspace{0.5em}
p_4\rightarrow p_4 \cos{\alpha}+p_5\sin{\alpha}
+\frac{2N_{c}e^2}{3\pi^{2}f_{\pi}^3} 
               (\bf B\cdot A^{\rm L}\rm)
(\theta_4 \cos{\alpha}+\theta_5 \sin{\alpha})
(\theta_3+\frac{1}{\sqrt{3}}\theta_8)\;\;,\\ 
\vspace{0.5em}
p_5 \rightarrow p_5 \cos{\alpha}-p_4\sin{\alpha}
+\frac{2N_{c}e^2}{3\pi^{2}f_{\pi}^3} 
               (\bf B\cdot A^{\rm L}\rm) 
(\theta_5 \cos{\alpha}-\theta_4 \sin{\alpha})
(\theta_3+\frac{1}{\sqrt{3}}\theta_8)\;\;,\\ 
\vspace{0.5em}
  p_8 \rightarrow p_8 +\frac{N_{c}e^2}{3\sqrt{3}\pi^{2}f_{\pi}^3}
               (\bf B\cdot A^{\rm L}\rm)(\theta_{1}^2+
                 \theta_{2}^2+\theta_{4}^2+
                 \theta_{5}^2)\;\; . \\
 \end{array}
\end{equation}
transform the fields $\theta_i\; (i=1,2,4,5)\;,\; p_i  
\;(i=1,2,3,4,5,8)$ and
$\mbox{\boldmath $\pi$}^T$ into BRST and $\sigma$ closed.
( The same notation as in the previous case is kept ).
\begin{displaymath}
  sA^{L'}\rm=-C\;\; , \;\; \hspace{1em}
      sC=0\;\; , \;\;
\end{displaymath}
\begin{displaymath}
s{\cal P}=\pi^{L'}-\rho^{(2)}-\rho^{(3)}\;\; , \;\;
\hspace{1em} 
      s\pi^{L'}=0\;\; , \;\;
\end{displaymath}
\begin{displaymath}
  sA_{0}=i\bar{{\cal P}}\;\; , \;\; \hspace{1em}
      s\bar{{\cal P}}=0\;\; , \;\;
\end{displaymath}
\begin{equation}
  s\bar{C}=-i\pi_{0}\;\; , \;\; \hspace{1em}
      s\pi_{0}=0\;\; , \;\;
\end{equation}
\begin{displaymath}
  s\theta_{\it a}=0\;\; , \;\; \hspace{1em}
      sp_{\it a}=0\;\; , \;\; \hspace{1em}
\it a=\rm 1,2,...,8
\end{displaymath}
\begin{displaymath}
  s\bf A^{\rm T}\rm=0\;\; , \;\; \hspace{1em}
      s\mbox{\boldmath $\pi$}^{T}=0\;\; , \;\;
\end{displaymath}
and 
\begin{displaymath}
  \sigma(-C)=A^{L'}\;\; , \;\; \hspace{1em}
      \sigma A^{L'}=0\;\; , \;\;
\end{displaymath}
\begin{displaymath}
\sigma \pi^{L'}={\cal P}\;\; ,
\;\; \hspace{1em} 
      \sigma {\cal P}=0\;\; , \;\;
\end{displaymath}
\begin{displaymath}
  \sigma(i\bar{{\cal P}})=A_{0}\;\; , \;\; \hspace{1em}
      \sigma A_{0}=0\;\; , \;\;
\end{displaymath}
\begin{equation}
  \sigma(-i\pi_{0})=\bar{C}\;\; , \;\; \hspace{1em}
      \sigma(\bar{C})=0\;\; , \;\;
\end{equation}
\begin{displaymath}
  \sigma\theta_{\it a}=0\;\; , \;\; \hspace{1em}
      \sigma p_{\it a}=0\;\; , \;\; \hspace{1em}
\it a=\rm 1,2,...,8
\end{displaymath}
\begin{displaymath}
  \sigma \bf A^{\rm T}\rm=0\;\; , \;\; \hspace{1em}
      \sigma \mbox{\boldmath $\pi$}^{T}=0\;\; . 
\end{displaymath}
Now we proceed as in the previous case and solve for 
C, $\bar{{\cal P}}$ , 
$ \mbox{\boldmath $\pi$}^L $ and $\pi_0$ from (17) 
\begin{equation}
C=-sA^{L'}\;\; , \;\;
\hspace{1em} 
\bar{{\cal P}}=-isA_{0} \;\; , \;\; \hspace{1em}
\mbox{\boldmath $\pi$}^L
=\frac{\nabla}{\nabla^2}(s{\cal P}+\rho^{(2)}+\rho^{(3)})\;\; , 
\;\; \hspace{1em}
\pi_0=is\bar{C}\;\; .
\end{equation}
After performing the transformations (16) in (13) and substituting 
the expression of C, $\bar{{\cal P}}$ , 
$ \mbox{\boldmath $\pi$}^L $ and $\pi_0$ from (19) we obtain the 
following expression for $ S_{BFV} $
\begin{equation}
S_{BFV}\rightarrow\int d^4 x[-\mbox{\boldmath $\pi$}^T
\cdot\dot{\bf A}^{\rm T}+ 
             \rm p_{\it a}\dot{\theta_{\it a}} - H_{C}^{(2)}
-H_{C}^{(3)}+s(F^{(2)}+F^{(3)})]
             +\int dt [\Psi,Q]
\end{equation}
where $ H_{C}^{(2)}$ and $F^{(2)}$ are given in (10),
\begin{eqnarray}
   H_{C}^{(3)}&\!\!\!\!=&\!\!\!\!-\frac{N_{c}e}{3\pi^{2}f_{\pi}^3}
         [\mbox{\boldmath $\pi$}^{\rm T}\times\bf \nabla \rm (\theta_3 + 
              \frac{1}{\sqrt{3}} \theta_8)-
              (p_3+\frac{1}{\sqrt{3}}p_8) \bf B\rm)]
              \cdot(\theta_1\bf \nabla \rm \theta_2-\theta_2\bf \nabla
              \rm\theta_1+\theta_4\bf \nabla \rm \theta_5-
              \theta_5\bf \nabla \rm\theta_4) \nonumber \\
      &\!\!\!\!-&\!\!\!\!\frac{N_{c}e}{\sqrt{3}\pi^{2}f_{\pi}^3}
            (\mbox{\boldmath $\pi$}^{\rm T}
\times\bf \nabla \rm \theta_8-p_8 \bf B\rm)
            \cdot(\theta_7\bf \nabla \rm \theta_6-\theta_6\bf \nabla
            \rm\theta_7) \nonumber \\
      &\!\!\!\!+&\!\!\!\! \frac{N_{c}e^2}{9\pi^{2}f_{\pi}^3}
              (\mbox{\boldmath $\pi$}^{\rm T}\cdot \bf B\rm)
              \{ [4(\theta_{1}^2+\theta_{2}^2)+
                5(\theta_{4}^2+\theta_{5}^2)] \theta_3 \nonumber \\ 
      &\!\!\!\!+&\!\!\!\!
             \sqrt{3}[2(\theta_{1}^2+\theta_{2}^2)+
             \theta_{4}^2+\theta_{5}^2]\theta_8 +
             2[(\theta_1 \theta_5 - \theta_2 \theta_4)\theta_7 + 
             (\theta_1 \theta_4 + \theta_2 \theta_5)\theta_6] \}
                    \nonumber \\
      &\!\!\!\!-&\!\!\!\! \frac{N_{c}e^2}{3\pi^{2}f_{\pi}^3}
          (\mbox{\boldmath $\pi$}^{\rm T}
\times \bf A^{\rm T}\rm) \cdot [\bf \nabla
          \rm(\theta_{1}^2+\theta_{2}^2+\theta_{4}^2+\theta_{5}^2)] 
          (\theta_3+\frac{1}{\sqrt{3}}\theta_8)\nonumber \\
      &\!\!\!\!-&\!\!\!\! \frac{2N_{c}e^2}{3\pi^{2}f_{\pi}^3}
         (\bf A^{\rm T}\cdot B\rm)(p_1\theta_1+p_2\theta_2+p_4\theta_4+ 
         p_5\theta_5)(\theta_3+\frac{1}{\sqrt{3}}\theta_8)\nonumber \\
      &\!\!\!\!-&\!\!\!\! \frac{N_{c}e}{3\pi^{2}f_{\pi}^3}
        [\bf B\cdot\bf \nabla \rm(\theta_3+\frac{1}{\sqrt{3}}\theta_8)]
        (p_2\theta_1-p_1\theta_2+p_5\theta_4-p_4\theta_5)\nonumber \\
      &\!\!\!\!-&\!\!\!\! \frac{N_{c}e}{\sqrt{3}\pi^{2}f_{\pi}^3}
           (\bf B\cdot\bf \nabla \rm \theta_8)
           (p_6\theta_7-p_7\theta_6) \; \; .\nonumber \\ 
\nonumber 
\end{eqnarray}
and
\begin{eqnarray}
   F^{(3)}&\!\!\!\!=&\!\!\!\!{\cal P}\frac{1}{\nabla^2}\rho^{(3)}
+\frac{N_{c}e}{3\pi^{2}f_{\pi}^3}
   [(\frac{\nabla}{\nabla^2}{\cal P})\times \bf \nabla \rm 
(\theta_3+\frac{1}{\sqrt{3}}\theta_8)]
  \cdot(\theta_1 \nabla\theta_2-\theta_2\nabla\theta_1
+\theta_4 \nabla\theta_5-\theta_5\nabla\theta_4)\nonumber \\
       &\!\!\!\!-&\!\!\!\! \frac{N_{c}e^2}{9\pi^{2}f_{\pi}^3}
              (\frac{\nabla}{\nabla^2}\cal{P})\cdot\bf B\rm
              \{ [4(\theta_{1}^2+\theta_{2}^2)+
                5(\theta_{4}^2+\theta_{5}^2)] \theta_3 \nonumber \\ 
      &\!\!\!\!+&\!\!\!\!
             \sqrt{3}[2(\theta_{1}^2+\theta_{2}^2)+
             \theta_{4}^2+\theta_{5}^2]\theta_8 +
             2[(\theta_1 \theta_5 - \theta_2 \theta_4)\theta_7 + 
             (\theta_1 \theta_4 + \theta_2 \theta_5)\theta_6] \}
                    \nonumber \\
&\!\!\!\!+&\!\!\!\!
              \frac{N_{c}e}{\sqrt{3}\pi^{2}f_{\pi}^3}
       [(\frac{\nabla}{\nabla^2}{\cal P})\times \nabla\theta_8]\cdot
(\theta_7 \nabla\theta_6-\theta_6 \nabla\theta_7)\nonumber\\
&\!\!\!\!+&\!\!\!\!
              \frac{N_{c}e^2}{3\pi^{2}f_{\pi}^3}
       [(\frac{\nabla}{\nabla^2}{\cal P})\times \bf A^{\rm T}\rm]\cdot
[\nabla(\theta_{1}^2+\theta_{2}^2+\theta_{4}^2+\theta_{5}^2)]
(\theta_3+\frac{1}{\sqrt{3}}\theta_8) \;\; . \nonumber  
\end{eqnarray}
Finally by fixing the gauge fermion as follows
\begin{displaymath}
  \Psi=-\int d^3 x (F^{(2)}+F^{(3)})
\end{displaymath}
we end up with a Coulomb-gauge expression for the effective
action. In \cite{PP2} in the context of the FJ formalism we 
solved the constraint 
$\rho^{(2)}+\rho^{(3)}-\bf\nabla\mbox{\boldmath $\pi$}\rm=0$ 
for $ \mbox{\boldmath $\pi$}^L $ and we substituted into the 
expression (12) for the effective Lagrangian density. The resulting 
uncanonical expression was diagonalized by performing the field 
transformations (16). So it shown that also in this case the field 
transformations that transform the fields of the model into BRST 
and $\sigma$ closed are the Darboux's transformations of the FJ 
approach.

%
%%%%%%%%%%%%%%%%%%%%%%%%%%%%%%%%%%%%%%%%%%%%%%%%%%%%%%%%%%%%%%%%%
%
\section{Conclusion}
\label{Conc}
The four dimensional SU(3) WZW model coupled to electromagnetism 
was treated in the context of the BFV formalism for constraint systems. 
Comparison was made with the FJ approach and common features were 
stressed. It was shown that the field redefinitions that transform 
the fields of the model into BRST and $\sigma$ closed are actually 
the Darboux's transformations of the FJ approach.\\
\vspace{1em}
\\
We wish to thank Dr. Kostas Skenderis for useful discussions.
  
%
%%%%%%%%%%%%%%%%%%%%%%%%%%%%%%%%%%%%%%%%%%%%%%%%%%%%%%%%%%%%%%%%%
%
\section{Appendix}
\label{app}

Our metric is $g_{\mu \nu}=diag(1,-1,-1,-1) \;\;$.
We choose $ e>0 $. We define $ \epsilon^{0123}=1$ . 
By $\mbox{\boldmath $\pi$}$ we
denote the electric field $\bf E\rm$ so that
$(\pi_{\mu},A^{\mu}) \;\; \mu=0,1,2,3 $ is a canonical pair. We
made use of the following
Poisson brackets
\begin{eqnarray}
  [A^{\mu}(\bf x,\rm t),\pi^{\nu}(\bf y,\rm t)] &\!\!\!\!=&\!\!\!\! 
  g^{\mu\nu} \delta(\bf x-y \rm) \; \; ,\nonumber \\
\vspace{1em}
  [\theta_{\it a}(\bf x,\rm t),p_{\it b}(\bf y,\rm t)] 
&\!\!\!\!=&\!\!\!\! 
   \delta_{\it ab} \delta(\bf x-y \rm) \; \; ,\nonumber \\
\vspace{1em}
  [C(\bf x,\rm t),{\cal P}(\bf y,\rm t)] &\!\!\!\!=&\!\!\!\! 
  - \delta(\bf x-y \rm) \; \; ,\nonumber \\
\vspace{1em}
[\bar{C}(\bf x,\rm t),\bar{{\cal P}}(\bf y,\rm t)] &\!\!\!\!=&\!\!\!\! 
  - \delta(\bf x-y \rm) \; \; . \nonumber 
\end{eqnarray}
The Grassmann parities of the fields are given by $\;\;\;$
$\epsilon_{A_\mu}=\epsilon_{\pi_\mu}=\epsilon_{\theta_{\it a}}
=\epsilon_{p_{\it a}}=0\;\; ,\;\; 
\epsilon_C=\epsilon_{\cal P}
=\epsilon_{\bar{C}}=\epsilon_{\bar{{\cal P}}}=1$
and their ghost number 
$gh(C)=-gh({\cal P})=1\;\;,\;\; gh(\bar{C})=-gh(\bar{{\cal
P}})=-1 \;\;,\;\;
gh(A_{\mu})=gh(\pi_{\mu})=gh(\theta_{\it a})=gh(p_{\it a})=0$.

%%%%%%%%%%%%%%%%%%%%%%%%%%%%%%%%%%%%%%%%%%%%%%%%%%%%%%%%%%%%%%%%%%%%%%%%

\end{document}